\documentclass[aps,prl]{revtex4}
\begin{document}

\title{SETI and muon collider}

\author{ Z.~K.~Silagadze}
\affiliation{
Budker Institute of Nuclear Physics and Novosibirsk State
University, 630 090, Novosibirsk, Russia }
%\date{}

\begin{abstract}
Intense neutrino beams that accompany muon colliders can be used for 
interstellar communications. The presence of multi-TeV extraterrestrial muon 
collider at several light-years distance can be detected after one year run 
of IceCube type neutrino telescopes, if the neutrino beam is directed towards 
the Earth. This opens a new avenue in SETI: search for extraterrestrial muon 
colliders. 
\end{abstract}
\maketitle

Are we alone in the immensely large universe? This is one of fundamental
questions steering the interest of broad public to SETI -- the Search for 
Extra-Terrestrial Intelligence. ``It is to everyone's benefit to nurture 
this interest in the real science of SETI rather than in the pseudoscience 
that preys on the public's credulity'' (Tarter, 2001). The theme of 
extraterrestrial creatures was always popular in human history and still 
abounds in popular culture. However, the real scientific SETI begins from the 
paper of Cocconi and Morrison some 50 years ago (Cocconi and Morrison, 1959), 
followed by the Project Ozma (Drake, 1965), the first dedicated search of 
extraterrestrial radio signals from two nearby Sun-like stars. Ever Since it 
was usually assumed that the centimeter wavelength electromagnetic signals 
are the best choice for interstellar communications. Here we question this 
old wisdom and argue that the muon collider, certainly in reach of modern day 
technology (Ankenbrandt {\it et al.}, 1999), provides a far more unique 
marker of civilizations like our own [type I in Kardashev's classification 
(Kardashev, 1964)]. Muon colliders are accompanied by a very intense and 
collimated high-energy neutrino beam which can be readily detected even at 
astronomical distances.

Muon collider was first suggested by Budker forty years ago (Budker, 1970). 
Ionization cooling, the idea that dates back to O'Neill (O'Neill, 1956), 
provides the possibility to make very bright muon beams (Skrinsky and 
Parkhomchuk, 1981). Muons are unstable particles and their decays produce 
neutrinos. Therefore, high-luminosity muon collider with long straight 
sections is also a neutrino factory producing the thin pencil beams of 
neutrinos (Geer, 1999). The expected neutrino intensities are so huge that 
even constitute a considerable radiation hazard in the neighborhood of the 
collider (King, 2000). Nevertheless, the present day technology is mature 
enough to make the construction of muon collider and hence neutrino factory 
quite realistic (Ankenbrandt {\it et al.}, 1999; Alsharoa {\it et al.}, 
2003). We may wonder whether extraterrestrial civilizations also built muon 
colliders and are illuminating us by accompanying neutrino beams. Can we 
detect these neutrinos from the alleged extraterrestrial muon colliders?

Due to relativistic kinematics, all neutrinos emitted by an 
ultra-relativis\-tic muon in the forward hemisphere in the muon rest frame 
will be boosted,  in  the laboratory frame, into a very narrow cone with an 
opening half-angle,
$$\theta\approx \frac{1}{\gamma}\approx \frac{10^{-4}}{E_\mu[TeV]},$$
where $\gamma$ is the relativistic boost factor of the muon and $E_\mu$ is
its energy.

Therefore, $E_\mu=200~\mathrm{TeV}$ extraterrestrial muon collider operating
at the $L=20$ light-years distance will illuminate with neutrinos a disk of
radius $R\approx L\theta\approx 10^8~\mathrm{km}$, which is somewhat smaller 
than the Earth's orbital radius. The neutrino flux on the Earth, assuming the 
Earth is inside of the neutrino disk, will be $\Phi_\nu\approx 10^5~\mathrm
{year}^{-1}~\mathrm{km}^{-2}$, if the neutrino beam intensity at the muon 
collider is $N_\nu=3\times 10^{21}~\mathrm{year}^{-1}$.

The main difficulty in neutrino detection is that neutrinos are very weakly
interacting elusive particles. One of methods of high-energy neutrino 
detection is to look for muons generated in charged-current interactions 
of neutrinos in the rock below the detector (Gaisser {\it et al.}, 1995) The 
muon should be generated within the muon range in the rock (about 
one kilometer for TeV muons) to reach the detector and produce observable 
signal through the Cherenkov radiation. The probability that a neutrino of 
energy $E_\nu$ will produce a muon within the muon range from the detector is 
approximately  $P_{\nu\to\mu}=1.7\times 10^{-6}E_\nu^{0.8}$ for multi-TeV 
neutrinos (Gaisser {\it et al.}, 1995; Halzen and Hooper, 2002). For 
$E_\nu=100~\mathrm{TeV}$ this gives $P_{\nu\to\mu}\approx  7\times 10^{-5}$.

The similar conclusion $P_{\nu\to\mu}\approx 10^{-4}$ can be reached from 
estimates of the probability of neutrino interaction in the effective 
detector volume, after penetrating through Earth from the gamma-ray burst in 
the northern hemisphere, in a km deep under-ice detector at the South Pole 
(Razzaque {\it et al.}, 2004). 

Therefore, for $S=1~\mathrm{km}^2$ area neutrino detectors, such as IceCube at
the South Pole (Ahrens et al., 2004) the expected rate of neutrino events 
from the hypothetical extraterrestrial muon collider is 
\begin{equation}
R=\Phi_\nu S P_{\nu\to\mu}\approx 7-10 ~\mathrm{year}^{-1}.
\label{eq1}
\end{equation}

Cosmic-ray induced background for IceCube is about 0.08 neutrino 
events with $E_\nu> 10~\mathrm{TeV}$ per year per square degree (Dermer, 
2007). In light of IceCube's very good angular resolution [better than 
$1^\circ$ (Ahrens {\it et al.}, 2004)], we conclude that detection of 
point-like  multi-TeV neutrino sources is essentially background free for 
such type of neutrino detectors and, therefore, (\ref{eq1}) constitutes a 
significant signal allowing to detect the presence of extraterrestrial muon 
collider at 20 light-years distance after one year run.

Note that the parameters of the muon collider we have assumed ($E_\mu=200~
\mathrm{TeV}$, $N_\nu=3\times 10^{21}~\mathrm{year}^{-1}$), although 
challenging for modern-day technology, are likely to be within its reach, 
at least for a single-pass muon colliders (Zimmermann, 2000). Therefore, 
(\ref{eq1}) should be considered as a lower bound for advanced civilizations. 
For example, a futuristic $10^3~\mathrm{TeV}$ muon collider was suggested  
(Sugawara {\it et al.}, 2003) to use the accompanying ultra high-energy 
neutrino beam for destruction of terrorist's concealed nuclear warheads. We 
hope that advanced civilizations capable to develop the necessary technology 
are already free from such nasty problems. However, we may imaging various 
peaceful applications of the high-energy neutrino beams, for example, for the 
study of the inner structure of the host planet (De Rujula {\it et al.}, 
1983). 

There have been proposals to use collimated neutrino beams for 
telecommunications (Saenz {\it et al.}, 1977; Ueberall {\it et al.}, 1979), 
including even interstellar communications  (Subotowicz, 1979; Pasachoff and 
Kutner, 1979). However, only now, on the eve of muon collider era, this 
fantastic idea acquires a realistic shape.

It is clear that practical realization of interstellar neutrino communications
requires higher level of technology than our civilization now possesses. It 
was suggested that advanced civilizations may deliberately choose the neutrino 
channel for interstellar and intergalactic communications to shutout very 
young and not mature emergent civilizations like our own from the 
conversation (Subotowicz, 1979).

Intergalactic neutrino communications will require much higher neutrino
energies and intensities. Maybe type III civilizations (which have captured 
the power of an entire host galaxy) can produce and control neutrino beams
even beyond the so called  Greisen-Zatseptin-Kuzmin limit of about 
$10^{19}~\mathrm{eV}$. Interestingly, Askaryan effect (Gorham {\it et al.},
2007) allows to develop large-scale detectors to detect such ultra 
high-energy neutrinos through the coherent Cerenkov radio signal created by 
an neutrino initiated electromagnetic shower in a salt dome. Hopefully we 
will soon have an operating detector of this type (Reil, 2006).

Neutrino SETI was also proposed earlier with somewhat different perspective 
(Learned {\it et al.}, 1994). It was suggested that type II (which have 
captured all of the power from their host star) and type III civilizations, 
spread throughout the Galaxy, may require interstellar time standards to 
synchronize their clocks. It is argued that mono-energetic $45.6~\mathrm{GeV}$ 
neutrino pulses from the $Z^0\to\nu\bar\nu$ decays produced in a futuristic 
dedicated electron-positron collider of huge luminosity may provide such 
standards. If there is an extraterrestrial civilization of this type nearer 
than about $1~\mathrm{kpc}$ using this synchronization method, the associated 
neutrinos can be detected by terrestrial neutrino telescopes with an effective 
volume of the order of $\mathrm{km}^3$ of watter (Learned {\it et al.}, 1994).

An appealing feature of the neutrino SETI is that it does not require any 
special efforts, in contrast to the radio SETI, and can be conducted in a 
background regime as a by product of the conventional neutrino astrophysics. 
There are several neutrino telescopes under construction world wide that will
allow neutrino detection in a broad energy range. We just should have in mind 
that some high-energy neutrino signals which will be detected by these devices 
might have artificial origin.       

We conclude that at the jubilee of the SETI proposal by Cocconi and Morrison 
it`s just the time to search for neutrino signals from extraterrestrial muon 
colliders. What is the probability of success?  ``The probability of success 
is difficult to estimate: but if we never search, the chance of success is 
zero'' (Cocconi and Morrison, 1959).

\section*{Acknowledgments}
The work is supported in part by grants Sci.School-905.2006.2 and 
RFBR 06-02-16192-a.

%\begin{thebibliography}{}

\section*{References}
\begin{itemize}
\item[]{}
Ahrens, J. {\it et al.}  [IceCube Collaboration] (2004) Sensitivity of the 
IceCube detector to astrophysical sources of high energy muon neutrinos.
{\it Astropart.\ Phys.\ }  {\bf 20}, 507-532 [arXiv:astro-ph/0305196].
%%CITATION = APHYE,20,507;%%

\item[]{}
Alsharoa, M. M. {\it et al.}  [Muon Collider/Neutrino Factory
Collaboration] (2003) Recent progress in neutrino factory and muon collider 
research within the Muon collaboration.
{\it  Phys.\ Rev.\ ST Accel.\ Beams} {\bf 6}, 081001 (2003)
[arXiv:hep-ex/0207031].
%%CITATION = PRSTA,6,081001;%%
 
\item[]{}
Ankenbrandt, C. M. {\it et al.} (1999)
Status of muon collider research and development and future plans.
{\it Phys.\ Rev.\ ST Accel.\ Beams} {\bf 2}, 081001 [arXiv:physics/9901022].
%%CITATION = PRSTA,2,081001;%%
 
\item[]{}
Budker, G. I. (1970) Accelerators and Colliding Beams (in Russian). 
In Proceedings of the 7th International Conf. on High Energy Accelerators, 
Yerevan, 1969, Academy of Sciences of Armenia, Yerevan, p. 33.
 
\item[]{}
Cocconi, G. and Morrison, P. (1959) Searching for interstellar communications. 
{\it Nature} {\bf 184}, 844-846.
%%CITATION = NATUA,184,844;%%
 
\item[]{}
Dermer, C. D. (2007) Best-bet astrophysical neutrino sources.
{\it J.\ Phys.\ Conf.\ Ser.\ }  {\bf 60}, 8-13 [arXiv:astro-ph/0611191].
%%CITATION = 00462,60,8;%%
 
\item[]{}
De Rujula, A. Glashow, S. L. Wilson, R. R. and Charpak, G. (1983)
Neutrino Exploration Of The Earth.
{\it  Phys.\ Rept.\ }  {\bf 99}, 341-396.
%%CITATION = PRPLC,99,341;%%
 
\item[]{}
Drake, F. D. (1965) Radio Search for Extraterrestrial Life. In Current Aspects
of Exobiology, edited by Mamikunian, G. and  Briggs, M. H., Pergamon, 
New York, pp. 323-345.
 
\item[]{}
Gaisser, T. K. Halzen, F. and Stanev, T. (1995)
Particle Astrophysics With High-Energy Neutrinos.
{\it  Phys.\ Rept.\ }  {\bf 258}, 173-236 
[Erratum-ibid.\  {\bf 271}, 355-356 (1996)] [arXiv:hep-ph/9410384].
%%CITATION = PRPLC,258,173;%%
 
\item[]{}
Geer, S. (1998) Neutrino beams from muon storage rings: characteristics
and physics potential. {\it Phys.\ Rev.\  D} {\bf 57}, 6989-6997
[Erratum-ibid.\  {\it D} {\bf 59}, 039903 (1999)] [arXiv:hep-ph/9712290].
%%CITATION = PHRVA,D57,6989;%%
 
\item[]{}
Gorham, P. W.  {\it et al.}  [ANITA Collaboration] (2007)
Observations of the Askaryan effect in ice.
{\it Phys.\ Rev.\ Lett.\ }  {\bf 99}, 171101 [arXiv:hep-ex/0611008].
%%CITATION = PRLTA,99,171101;%%
 
\item[]{}
Halzen, F. and Hooper, D. (2002)
High-energy neutrino astronomy: The cosmic ray connection.
{\it Rept.\ Prog.\ Phys.\ } {\bf 65}, 1025-1078  [arXiv:astro-ph/0204527].
%%CITATION = RPPHA,65,1025;%%
 
\item[]{}
Kardashev, N. S. (1964) Transmission of information by extraterrestrial 
civilizations. Sov. Astron. {\bf 8}, 217-221.
%%CITATION = SAAJA,8,217;%%
 
\item[]{}
King, B. J. (2000) Neutrino radiation challenges and proposed solutions for 
many-TeV muon colliders. {\it  AIP Conf.\ Proc.\ }  {\bf 530}, 165-180 
[arXiv:hep-ex/0005006].
%%CITATION = APCPC,530,165;%%
 
\item[]{}
Learned, J. G. Pakvasa, S. Simmons, W.A. and Tata, X. (1994)
Timing data communication with neutrinos: A new approach to SETI.
{\it Q.\ J.\ Roy.\ Astron.\ Soc.\ }  {\bf 35}, 321-329.
%%CITATION = QJRAA,35,321;%%
 
\item[]{}
Pasachoff, J. M. and Kutner, M. L. (1979) Neutrinos for Interstellar 
Communication. {\it Cosmic Search} {\bf 1N3}, 2-8.
 
\item[]{}
O'Neill, G. K. (1956) Storage-Ring Synchrotron: Device for High-Energy Physics 
Research. {\it Phys.\ Rev.\ } {\bf 102}, 1418-1419.
%%CITATION = PHRVA,102,1418;%%
 
\item[]{}
Razzaque, S. Meszaros, P. and Waxman, E. (2004)
Neutrino signatures of the supernova - gamma ray burst relationship.
{\it  Phys.\ Rev.\   D} {\bf 69}, 023001 [arXiv:astro-ph/0308239].
%%CITATION = PHRVA,D69,023001;%%
 
\item[]{}
Reil, K.  [SalSA Collaboration] (2006)
SalSA: A Teraton UHE neutrino detector.
{\it  AIP Conf.\ Proc.\ } {\bf 842}, 980-982.
%%CITATION = APCPC,842,980;%%

\item[]{}
Saenz, A. W. Ueberall, H. Kelly, F. J. Padgett, D. W. and Seeman, N. (1977) 
Telecommunications with Neutrino Beams. 
{\it Science} {\bf 198}, 295-297.
%%CITATION = SCIEA,198,295;%%

\item[]{}
Skrinsky, A. N.  and Parkhomchuk, V. V. (1981) Cooling Methods For Beams Of 
Charged Particles. {\it Sov.\ J.\ Part.\ Nucl.\ }  {\bf 12}, 223-247.
%%CITATION = FECAA,12,557;%%
 
\item[]{}
Subotowicz, M. (1979) Interstellar Communication by Neutrino Beams.
{\it Acta Astronautica} {\bf 6}, 213-220.
%%CITATION = AASIC,6,213;%%
 
\item[]{}
Sugawara, H. Hagura, H. and Sanami, T. (2003)
Destruction of nuclear bombs using ultrahigh-energy neutrino beam.
arXiv:hep-ph/0305062.
%%CITATION = HEP-PH/0305062;%% 
 
\item[]{}
Tarter, J. (2001) The Search for Extraterrestrial Intelligence.
{\it Annu. Rev. Astron. Astrophys.} {\bf 39}, 511-48.
%%CITATION = ARAAA,39,511;%%
 
\item[]{}
Ueberall, H. Kelly, F. J. and Saenz, A. W. (1979) Neutrino Beams: A New 
Concept in Telecommunications. {\it J.\ Wash.\ Acad.\ Sci.\ } {\bf 69}, 48-54.
 
\item[]{}
Zimmermann, F. (2000) Final focus challenges for muon colliders at highest 
energies. {\it  AIP Conf.\ Proc.\ }  {\bf 530}, 347-367.
%%CITATION = APCPC,530,347;%%

\end{itemize}
%\end{thebibliography}

\end{document}